\documentclass{article}

\usepackage[final]{neurips_2023}
\makeatletter
\renewcommand{\@noticestring}{} 
\makeatother

\usepackage[utf8]{inputenc} 
\usepackage[T1]{fontenc}    
\usepackage{hyperref}       
\usepackage{url}            
\usepackage{booktabs}       
\usepackage{amsfonts}       
\usepackage{nicefrac}       
\usepackage{microtype}      
\usepackage{xcolor}         
\usepackage{graphicx}
\usepackage{verbatim}

\usepackage{listings}

\usepackage{fancyhdr} 
\title{Biotic Browser: Applying StreamingLLM as a Persistent Web Browsing Co-Pilot}

\author{%
  Kevin F.~Dunnell\thanks{https://www.kevindunnell.com/} \\
  Media Lab\\
  Massachusetts Institute of Technology\\
  Cambridge, MA 02139 \\
  \texttt{dunnell@mit.edu} \\
  \And
  Andrew P.~Stoddard\thanks{https://www.apstodd.com/} \\
  Electrical Engineering \& Computer Science\\
  Massachusetts Institute of Technology\\
  Cambridge, MA 02139 \\
  \texttt{apstodd@mit.edu} \\
}

\begin{document}

\maketitle

\thispagestyle{fancy} 
\fancyfoot[L]{\textit{Written December 2023}} 

\begin{abstract}
This paper presents "Biotic Browser," an innovative AI assistant leveraging StreamingLLM to transform web navigation and task execution. Characterized by its ability to simulate the experience of a passenger in an autonomous vehicle, the Biotic Browser excels in managing extended interactions and complex, multi-step web-based tasks. It marks a significant advancement in AI technology, particularly in the realm of long-term context management, and offers promising applications for enhancing productivity and efficiency in both personal and professional settings.
\end{abstract}

\section{Introduction}

The recent explosion of large language models (LLMs) has created a surge in the development of new "intelligent" software applications, particularly in persistent personal assistants. While numerous efforts are being made to build such an assistant, a significant gap remains in addressing user experience. Many of these efforts overlook the intricacies of human-AI interaction, a crucial element for successfully adopting such technologies. The Biotic Browser is our response to this gap. The tool introduces a user experience akin to being a passenger in an autonomous vehicle. This novel approach marks a significant departure from conventional text-based AI user interfaces, where web browsing by an LLM is often hidden from the user. Biotic Browser allows users to engage with web navigation intuitively and transparently, thereby enhancing trust and ease of use. This design philosophy addresses the growing demand for AI systems that are efficient but also user-friendly and trustworthy, setting a new precedent in the field.

Despite the capabilities of state-of-the-art LLMs, they often fall short in tasks requiring the processing and retention of extended interaction sessions. This is particularly evident in applications demanding dynamic and continuous user interaction, such as complex web navigation and sophisticated task management. StreamingLLM [1] addresses this limitation, enabling the retention and processing of information over prolonged interactions. Biotic Browser, leveraging StreamingLLM, represents a significant advancement in this area. It offers an AI assistant adept at managing multi-step tasks and maintaining a comprehensive history of user interactions. Designed as an advanced web browser agent, the Biotic Browser can execute intricate plans and interact seamlessly with various web platforms. By combining the technological capabilities of StreamingLLM with a focus on continuous context management and user-centric design, the Biotic Browser not only enhances the practicality of AI applications in personal and professional settings but also showcases the potential of StreamingLLM technology in creating a more intuitive and capable digital assistant.

\section{Methodology}
\subsection{Biotic Browser + GPT-4 Vision}
Inspired by the release of GPT-4 Vision (GPT-4V) [2], we initially leveraged this technology in the development of Biotic Browser to create a solution that integrates both image and text processing for an AI-assisted web browsing experience. The system begins by capturing a user-defined goal, such as 'buy an iPhone.' With this goal, Biotic Browser takes a screenshot of the current webpage, performs preprocessing on the screenshot to identify and number interactive elements, and then passes the annotated images along with textual descriptions of the goal to GPT-4V. The model then returns a set of proposed actions in JSON format or a completion state of 'true' if the task is completed. This process continues in a loop until GPT-4V confirms that the goal has been achieved. An outline of this architecture is presented in Figure 1.

While innovative, this approach faced several limitations. First, in terms of data privacy, it required sharing screenshots of web browsing for inference. Secondly, we encountered a bottleneck in terms of speed due to API latency. Additionally, during extended browsing sessions, the system faced challenges with memory cache overflow. Another significant issue was balancing the execution of actions relevant to the current step while still aligning with the overarching long-term goal, a complex task for the AI to manage effectively.

We explored open-source multimodal (text + image) alternatives to GPT-4V, such as LLaVa [3,4] and miniGPT-4 [5]. However, neither of these options could caption webpage screenshots with the same fidelity as GPT-4V. While the release of GPT-4V was a key motivator in our initial development of Biotic Browser using a multimodal model, the lack of suitable open-source alternatives led us to reconsider our architecture and shift to using a text-only model.

\begin{figure}
  \centering
  \includegraphics[width=1\textwidth]{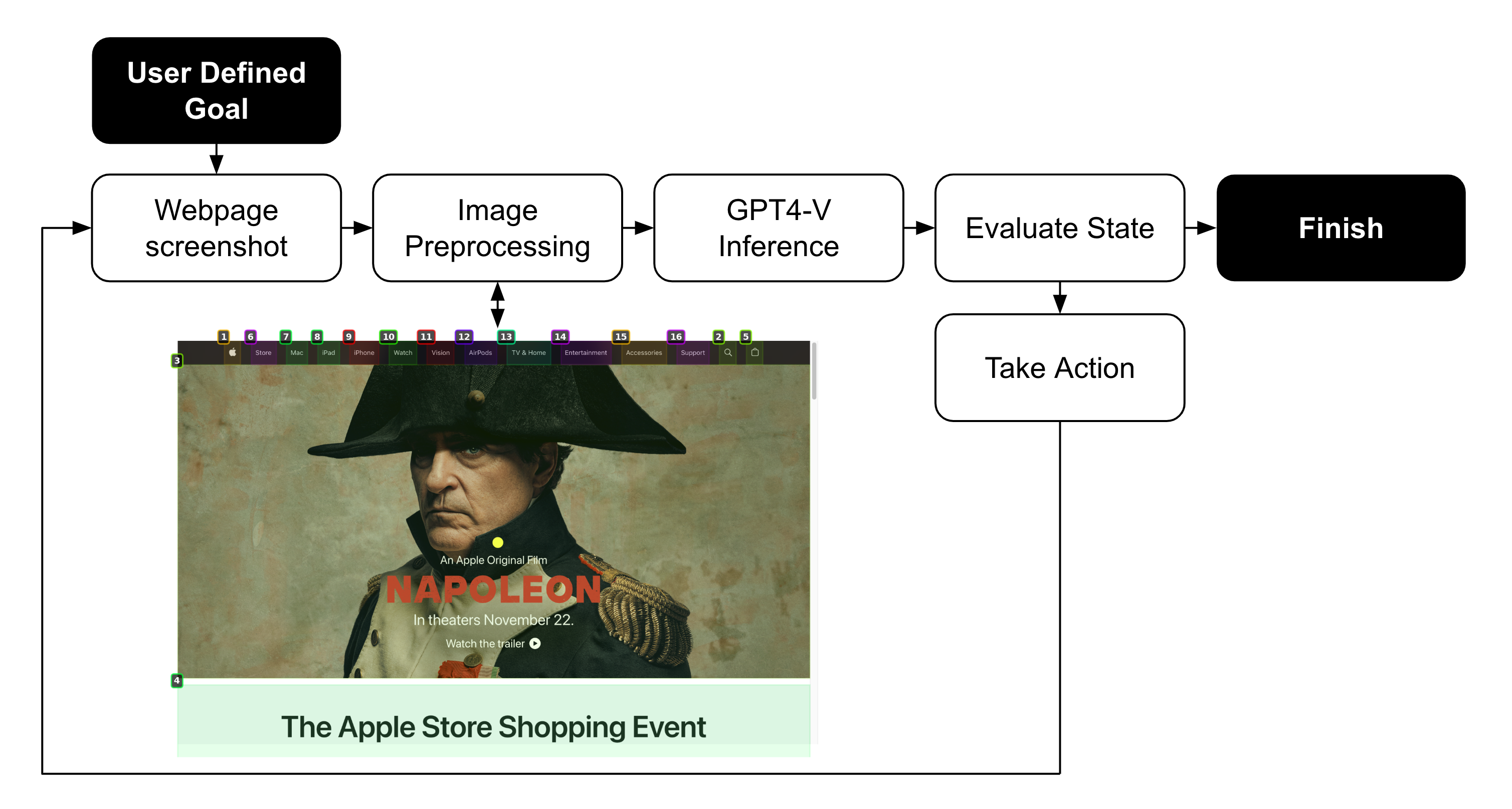}
  \caption{Biotic browser architecture with GPT-4V}
\end{figure}

\subsection{Biotic Browser + StreamingLLM}

In response to the limitations encountered with Biotic Browser + GPT-4V, we pivoted our approach to incorporate StreamingLLM. StreamingLLM's capability to manage and retain extended conversation histories and task lists proved to be a more fitting framework for the tool's objectives. This integration allowed Biotic Browser to process lengthy and complex task sequences without requiring visual data to be sent to OpenAI. As a result, this enhanced the system's data privacy considerably.

In this updated architecture, the system scrapes and processes the DOM to identify and tag interactable elements on the page instead of taking screenshots. These elements, along with the user-defined goals, are passed to StreamingLLM. The model then either confirms task completion or suggests the next action in JSON format. Figure 2 presents an outline of this new architecture, while Table 1 details the structure of the data input and output to the LLM. 

A significant addition in this iteration is the system's enhanced capability to request more information from the user if it encounters uncertainty or confusion in task progression. A more detailed view of the prompts sent to the LLM is available in Appendix A.

\begin{figure}
  \centering
  \includegraphics[width=1\textwidth]{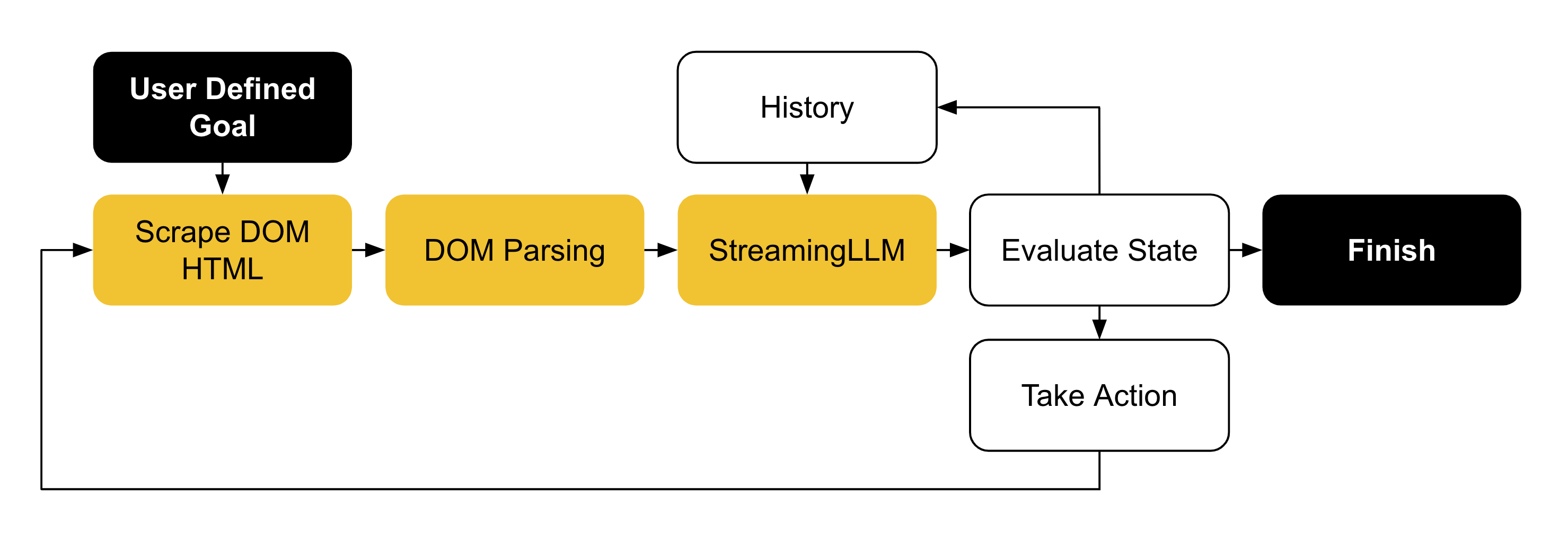}
  \caption{Biotic browser architecture with StreamingLLM}
\end{figure}

\begin{table}
  \caption{Inputs and Outputs Description}
  \label{inputs-outputs-table}
  \centering
  \begin{tabular}{lll}
    \toprule
    Category & Subcategory & Details \\
    \midrule
    {Inputs} & Overall Task & String definition of ultimate goal for the assistant \\
                            & Elements & JSON object representing elements on page \\
                            & Next Step & String description of next action to be taken \\
                            & History & String description of previous actions taken \\
                            & Clarifying Questions & Array of string questions asked previously \\
    \midrule
    {Outputs} & {Event List} & Click, Cursor Move, Scroll, and Text Input \\
                             & Next Step & String description of next action to be taken \\
                             & Is Complete & Boolean indicating if overall task is completed \\
                             & Questions & Array of string questions from system \\
                             & Actions & String of action taken at this step \\
    \bottomrule
  \end{tabular}
\end{table}

\subsection{User Experience Design}
Central to the development of the Biotic Browser is its innovative user interface design, which draws inspiration from the experience of being a passenger in an autonomous vehicle. This design paradigm represents a significant shift from traditional web navigation interfaces, focusing on creating an immersive, intuitive, and engaging user experience. The interface allows users to observe the browser's actions as it navigates through web pages, akin to watching an autonomous vehicle chart its course. This visualization of the browser's decision-making process is not only informative but also enhances the user's sense of control and trust in the system. Users are not passive observers; they can intervene and redirect the browser's actions at any point, similar to a driver taking control of a vehicle. This level of interaction and control is a response to the need for AI systems that are transparent and trustworthy, addressing common concerns about the opacity of AI decision-making processes. By combining these elements, Biotic Browser's user interface proposes a new standard in AI-driven web navigation, prioritizing user control, which we hope will help with the adoption of AI-assisted technology.

\section{Results}
The integration of StreamingLLM into Biotic Browser presented notable challenges, chiefly memory allocation constraints in local deployment. StreamingLLM's local performance was limited, producing approximately one token per minute. However, leveraging Google Colab for deployment significantly improved this speed. By manually transferring data between a local instance of Biotic Browser and StreamingLLM on Google Colab, we demonstrated the feasibility of this integration. Our success was short-lived due to a shortage of A100 GPUs, which were essential for our memory requirements, given the length of the prompts needed for StreamingLLM to produce the desired output (see Appendix A for details).

We significantly reconfigured Biotic Browser to operate effectively with language-based inputs, transitioning away from reliance on visual data processing. This shift involved parsing webpages' Document Object Model (DOM), moving from user screenshots, which raised privacy concerns, to a text-only approach. This adaptation not only addressed privacy issues but also enhanced Biotic Browser's integration capabilities with StreamingLLM.

While achieving a high-performing local deployment of StreamingLLM remains a goal, advancements in Biotic Browser in a text-centric format are promising. They underscore the potential of StreamingLLM's continuous context stream, suggesting more robust and efficient operation once local implementation hurdles are overcome.

Figure 3 presents a working demonstration of Biotic Browser, highlighting the user-defined input goal on the left and system feedback on the right. The red mouse cursor indicates AI control, allowing user observation and intervention.

\begin{figure}
  \centering
  \includegraphics[width=1\textwidth]{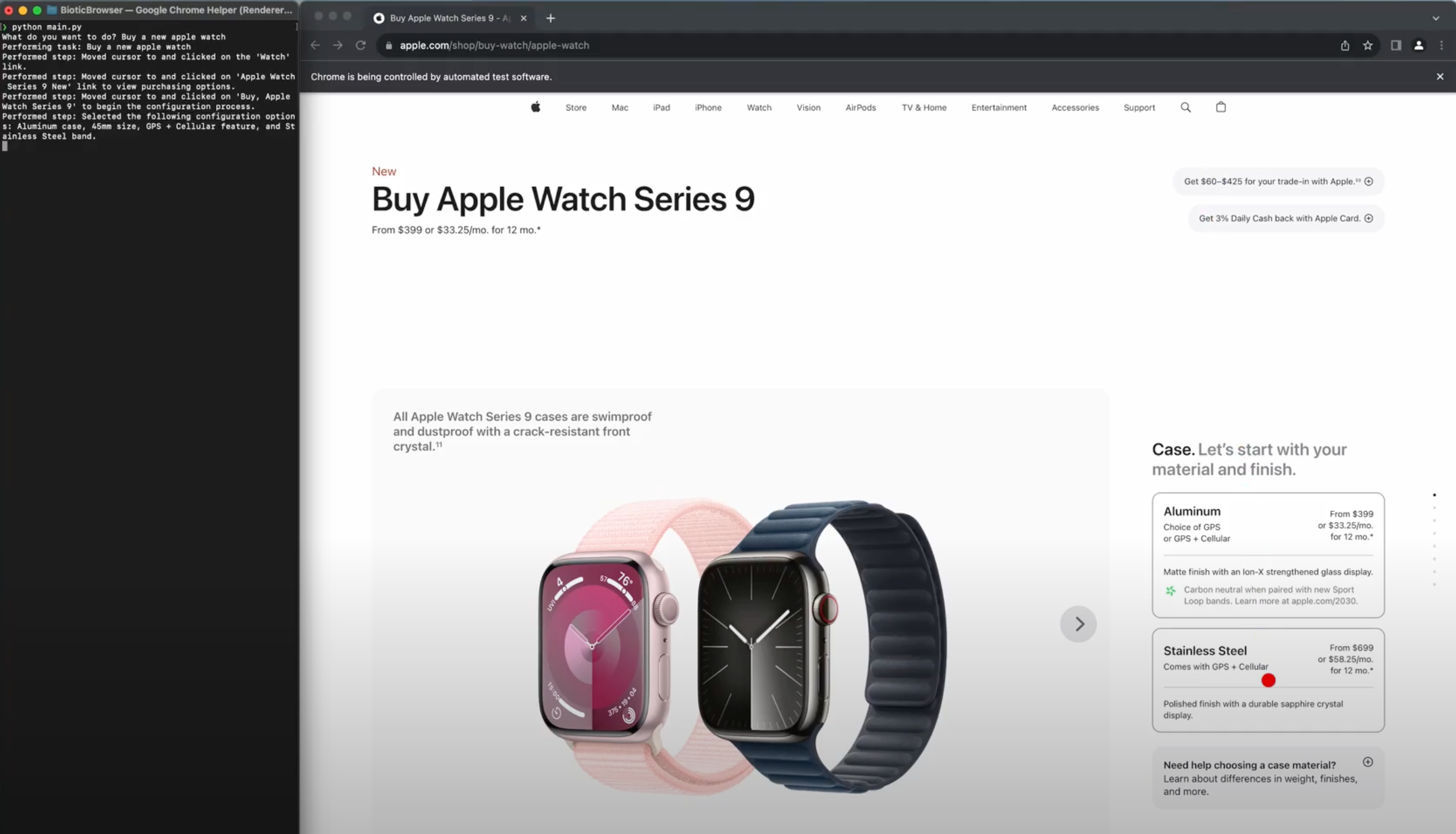}
  \caption{Biotic browser demo screenshot}
\end{figure}

\section{Discussion}

The development of Biotic Browser, integrated with StreamingLLM, marks a significant advancement in AI-assisted web navigation. This innovation fills a critical gap in current AI applications by seamlessly blending advanced language processing with an intuitive, user-focused web interface. The Biotic Browser's novel approach to web navigation parallels the experience of being a passenger in an autonomous vehicle, representing a shift in human-computer interaction.

Integrating StreamingLLM posed significant technical challenges, particularly in managing its local deployment due to resource requirements. These challenges reflect a broader issue in AI development: the need to balance the computational demands of advanced models with practical computing environments.

Shifting to a purely text-based approach while resolving operational and privacy issues introduced new complexities. This required rethinking the Biotic Browser's interaction with web content, moving from visual analysis to parsing the DOM of web pages.

The Biotic Browser's potential in personal and professional settings is immense. It could revolutionize task management and efficiency, from online shopping to multi-step project execution. This project contributes technically to AI and sets new standards for user-focused design, influencing future AI developments.

\section{Conclusion}
This paper has underscored the significant potential of Biotic Browser, an AI-assisted web browsing tool. Integrating StreamingLLM with a user-centric design, Biotic Browser offers an innovative approach to digital interaction. Its proficiency in managing complex tasks over extended periods demonstrates its capability to enhance productivity and efficiency in various settings. Biotic Browser represents a paradigm shift in AI-assisted web navigation and task execution. It paves the way for future developments in creating more intuitive and capable digital assistants.

\newpage
\section*{References}

\small

[1] Xiao, G., Tian, Y., Chen, B., Han, S., \ \& Lewis, M. (2023). Efficient Streaming Language Models with Attention Sinks. {\it arXiv}.

[2] OpenAI (2023). GPT-4V(ision) System Card. Retrieved from https://openai.com/index/gpt-4v-system-card/

[3] Liu, H., Li, C., Li, Y., \ \& Lee, Y. J. (2023). Improved Baselines with Visual Instruction Tuning. {\it arXiv}.

[4] Liu, H., Li, C., Wu, Q., \ \& Lee, Y. J. (2023). Visual Instruction Tuning. NeurIPS.

[5] Zhu, D., Chen, J., Shen, X., Li, X., \ \& Elhoseiny, M. (2023). MiniGPT-4: Enhancing Vision-Language Understanding with Advanced Large Language Models. {\it arXiv}.




\newpage 
\section*{Appendix A}

Biotic Browsing prompting instructions:

\begin{lstlisting}[language=Python]
prompt = """
    As an autonomous AI agent, your task is to navigate and interact with web pages, keeping track of
    significant actions and decisions made. Use this historical context to guide your current and
    future interactions towards completing the web task.

    Your input will be:
    - Task Description: A string with no formatting that describes the overall task to be completed. This is the most
    important information for you to understand, and will be provided at the beginning of each task. This should always
    be the overall goal.
    - Elements: A JSON object where each key corresponds to a unique identifier for an element visible
    on the current webpage, and the value is a dictionary containing information about the element such as 'tag_name',
    'accessible_name', 'aria_role', 'id', 'class', 'text', 'location', and 'size'.
    - Next Step: A string with no formatting that describes what you should do at the current step. If this is the
    first step, the next step will be blank, and you should perform the first action based on the task description.
    If the next step does not make sense based on the elements provided, you may ask clarifying questions or
    perform a different action that you think is more appropriate to achieve the task.
    - History: A string with no formatting that describes the actions you have taken so far. This will be blank
    for the first step, and will be updated after each step.
    - Clarifying Questions: An array of strings with no formatting that represent questions you have previously
    asked about the task or the provided text descriptions. These questions have been answered by a human and used
    to improve your understanding of the task.

    After processing the input, respond with a JSON object that strictly follows the defined structure below. The JSON object may
    not have escape characters and must be able to be parsed by json.loads().

    Output:
    - 'event_list': An array where each element is an object representing an interaction event. Each interaction
      event object must have a 'type' key that indicates the kind of event, such as 'click', 'cursor_move', 'scroll',
      or 'text_input'. If you want to click on an item or enter text, you must first move the cursor to the item.
      If the item is not in view, as given by the visible_in_viewport value, you must first scroll to the item. If
      a scroll event is required, you must first scroll to the item before moving the cursor to the item. You may
      not perform an action on an item that has not been moved to. The object structure for each type is:
      - For 'click' events: An object with 'type': 'click', and 'item': x, where x is an integer representing the
        item to interact with based on the text description.
      - For 'cursor_move' events: An object with 'type': 'cursor_move', and 'item': x, where x is an integer
        representing the item that the cursor should move to based on the text description.
      - For 'scroll' events: An object with 'type': 'scroll', and 'item': x, where x is an integer representing
        the item to scroll to based on the text description.
      - For 'text_input' events: An object with 'type': 'text_input', 'item': x, and 'text': 'string', where x is
        an integer representing the item to enter text into and where 'string' is the text to be entered based on
        the text description.
    - 'next_step': A string with no formatting that describes what you should do after the current set of events has
      been completed.
    - 'is_complete': A boolean value indicating whether the task has been fully completed.
    - 'questions': An array of strings with no formatting that represent questions you have about the task or the
      provided text descriptions. You may also ask questions if you need custom user input for a text field. These
      questions will be answered by a human and used to improve the AI's performance. You may leave this array empty
      if you have no questions, but always include the key. If you have questions, you may leave the event_list empty
      and the AI will not perform any actions on this page. However, you must still provide all of the other values.
      The next_step should describe what the user should do after the clarifying questions have been answered. The
      is_complete value should be false if the task is not fully complete on this page. The action value should
      indicate that you needed to ask clarifying questions, and what those questions were. You will not receive
      any additional instructions other than the clarifying questions.
    - 'action': A string with no formatting that describes the action you took at this step. This will be recorded
      in the history for future steps, so make sure it is accurate and will help you complete the task.

    Task Description:
    search for pizza

    Elements:
    {"0": {"tag_name": "a", "accesible_name": "About", "aria_role": "link", "id": "", "class": "MV3Tnb", "text": "About", "location": {"x": 21, "y": 17}, "size": {"height": 26, "width": 47}, "visible_in_viewport": true}, "1": {"tag_name": "a", "accesible_name": "Store", "aria_role": "link", "id": "", "class": "MV3Tnb", "text": "Store", "location": {"x": 78, "y": 17}, "size": {"height": 26, "width": 43}, "visible_in_viewport": true}, "2": {"tag_name": "a", "accesible_name": "Advertising", "aria_role": "link", "id": "", "class": "pHiOh", "text": "Advertising", "location": {"x": 175, "y": 1196}, "size": {"height": 46, "width": 99}, "visible_in_viewport": true}, "3": {"tag_name": "a", "accesible_name": "Business", "aria_role": "link", "id": "", "class": "pHiOh", "text": "Business", "location": {"x": 274, "y": 1196}, "size": {"height": 46, "width": 87}, "visible_in_viewport": true}, "4": {"tag_name": "a", "accesible_name": "How Search works", "aria_role": "link", "id": "", "class": "pHiOh", "text": "How Search works", "location": {"x": 361, "y": 1196}, "size": {"height": 46, "width": 147}, "visible_in_viewport": true}, "5": {"tag_name": "a", "accesible_name": "Our third decade of climate action: join us", "aria_role": "link", "id": "", "class": "pHiOh", "text": "Our third decade of climate action: join us", "location": {"x": 447, "y": 1149}, "size": {"height": 47, "width": 306}, "visible_in_viewport": true}, "6": {"tag_name": "a", "accesible_name": "Privacy", "aria_role": "link", "id": "", "class": "pHiOh", "text": "Privacy", "location": {"x": 801, "y": 1196}, "size": {"height": 46, "width": 76}, "visible_in_viewport": true}, "7": {"tag_name": "a", "accesible_name": "Terms", "aria_role": "link", "id": "", "class": "pHiOh", "text": "Terms", "location": {"x": 877, "y": 1196}, "size": {"height": 46, "width": 68}, "visible_in_viewport": true}, "8": {"tag_name": "a", "accesible_name": "Sign in", "aria_role": "link", "id": "", "class": "gb_za gb_jd gb_Ld gb_ie", "text": "Sign in", "location": {"x": 1086, "y": 12}, "size": {"height": 36, "width": 96}, "visible_in_viewport": true}, "9": {"tag_name": "input", "accesible_name": "Google Search", "aria_role": "button", "id": "", "class": "gNO89b", "text": "", "location": {"x": 459, "y": 453}, "size": {"height": 36, "width": 127}, "visible_in_viewport": true}, "10": {"tag_name": "input", "accesible_name": "I'm Feeling Lucky", "aria_role": "button", "id": "gbqfbb", "class": "", "text": "", "location": {"x": 598, "y": 453}, "size": {"height": 36, "width": 143}, "visible_in_viewport": true}}

    Next Step:


    History:
    No history yet.

    Clarifying Questions:
    No clarifying questions yet.
"""

prompt = prompt.replace('"', '').replace("'", '')
\end{lstlisting}

\end{document}